\documentclass{iau} 
\usepackage{graphicx}

\title[SWAG] 
{SWAG: Survey of Water and Ammonia in the Galactic Center}

\author[J\"urgen Ott et al.]   
{J\"urgen Ott$^1$,
David S. Meier$^2$, 
Nico Krieger$^3$,
Matthew Rickert$^4$,
\and the SWAG team}

\affiliation{$^1$ National Radio Astronomy Observatory, 
Socorro, NM, USA, email: {\tt jott@nrao.edu}
  \\[\affilskip]
$^2$ New Mexico Institute of Mining and Technology, 
Socorro, NM, USA 
\\ [\affilskip]
$^3$ Max-Planck-Institut f\"ur Astronomie, 
Heidelberg, Germany
\\ [\affilskip]
$^4$ Northwestern University, 
Evanston,  IL, USA\\ 
}

\pubyear{2016}
\volume{322}  
\jname{The Multi-Messenger Astrophysics of the Galactic Centre}
\editors{Roland Crocker, Steve Longmore, Geoff Bicknell, eds.}

\begin{document}

\maketitle

\begin{abstract}
SWAG (``Survey of Water and Ammonia in the Galactic Center'') is a
  multi-line interferometric survey toward the Center of the Milky
  Way conducted with the Australia Telescope Compact Array. The survey region spans the entire $\sim$400\,pc Central
  Molecular Zone and comprises $\sim$42 spectral lines at pc spatial and sub-km/s spectral resolution. In addition, we deeply map
  continuum intensity, spectral index, and polarization at the
  frequencies where synchrotron, free-free, and thermal dust sources
emit. The observed spectral lines include many transitions of
  ammonia, which we use to construct maps of molecular gas
  temperature, opacity and gas formation temperature (see poster by
  Nico Krieger et al., this volume). Water masers pinpoint the sites
  of active star formation and other lines are good tracers for
  density, radiation field, shocks, and ionization. This extremely rich
  survey forms a perfect basis to construct maps of the physical
  parameters of the gas in this extreme environment.

\end{abstract}

\firstsection 
\section{Introduction}

The Galactic Center, being the closest galactic nucleus at a distance
of only 8.5\,kpc, contains $\sim10$\% of the entire Galactic molecular
gas within a $\sim$500\,pc region known as the Central Molecular Zone
(CMZ; e.g. Morris et al. 1996; Oka et al. 1998; Jones et
al. 2012). This gas is subject to a unique and extreme environment
that is dominated by shocks from cloud-cloud collisions, large fluxes
of UV photons and cosmic rays, turbulence, as well as strong magnetic
and tidal fields. These conditions are reflected in extreme physical
parameters: large temperatures, densities, dispersions, and ionization
fractions of the gas, and in a peculiar chemistry that is enhanced in
abundances of complex molecules. Similar conditions are also found in
high-z galaxies at cosmic times when the bulk of stars were created
(e.g. Kruijssen \& Longmore 2013). To understand the astrophysics of
star formation processes in this environment, one must have accurate
measurements of these ISM conditions, as all of these physical
parameters determine the capability of the molecular gas to cool,
collapse, and eventually to form stars. Of particular importance is
the temperature of the gas, as it is a direct indicator of the
efficiencies of shielding and cooling.

\section{SWAG}
``The Survey of Water and Ammonia in the Galactic Center'' (SWAG;
survey page at: {\tt https://sites.google.com/site/atcaswag}) is an
approach to measure some of the most important quantities in the
CMZ. SWAG data consists of $\sim 42$ spectral lines in the 21-26\,GHz
range with 4\,GHz of continuum at  $\sim$0.8\,pc (20'') spatial and
  0.4\,km\,s$^{-1}$ spectral resolution. The data are a good compromise between spatial
resolution and surface brightness sensitivity, but the interferometric
observations do filter out the diffuse gas on the largest scales. The
maps therefore reveal the clumpy, filamentary structures of the CMZ in
great detail with little confusion.

\begin{figure}[h]
\begin{center}
 \includegraphics[width=\textwidth]{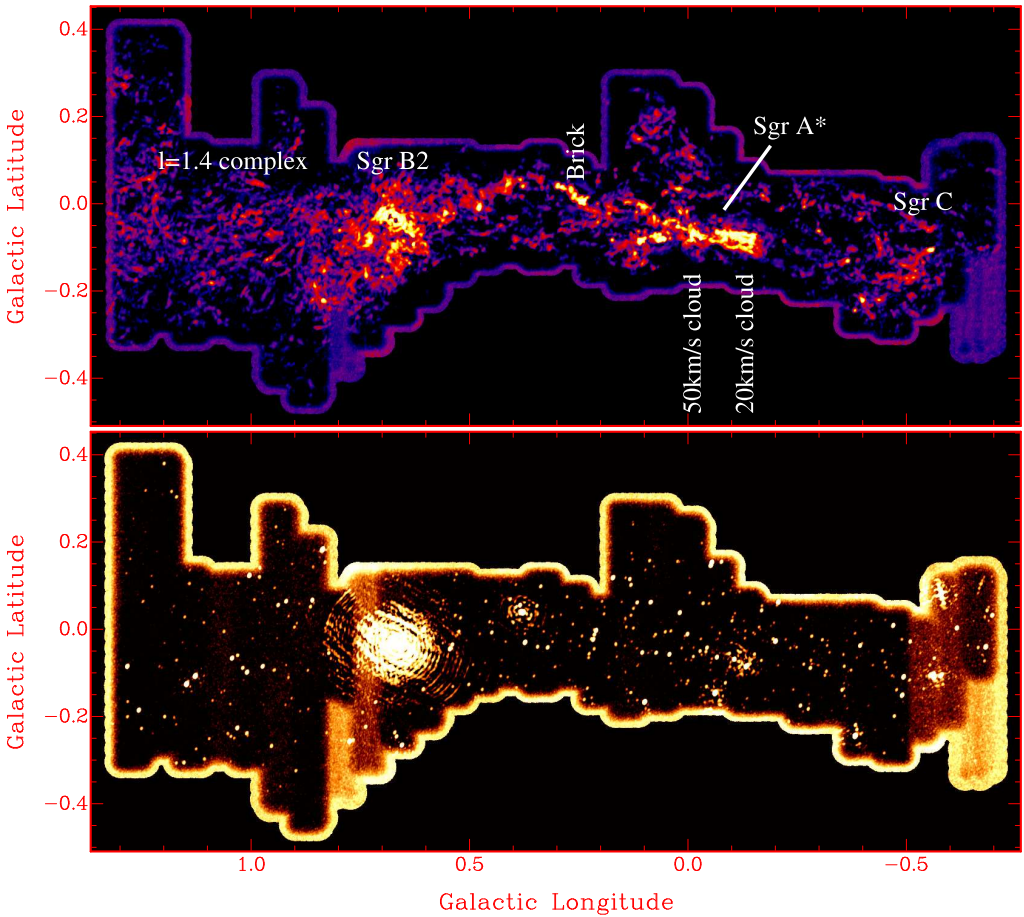} 
 \caption{{\it Top:} Ammonia (3,3) peak flux map from the first two
   years of SWAG. {\it Bottom:} 22\,GHz water maser
   peak flux map. Both maps cover $\sim 2/3$ of
   the final SWAG area.}
   \label{fig1}
\end{center}
\end{figure}

The SWAG lines include multiple ammonia transitions (Fig.\,\ref{fig1}
top), which are being used as a thermometer, structure, and dynamic
tracer for the molecular gas. Water masers (Fig.\,\ref{fig1} bottom)
are visible at sites where young stars currently form (in addition to
envelopes of evolved stars). We also obtain a rich dataset of density,
photo-dominated region and shock tracers (including c-C$_3$H$_2$,
HNCO, CH$_3$OH), radio recombination lines to map the ionized
components, and continuum maps to measure the strength, spectral
index, and polarization of the emission. Together, SWAG is a
comprehensive survey to couple the physical state (density,
temperature, ionization, shocks, magnetic field strength) of the
molecular gas with the dynamical structure of each cloud embedded in
the galactocentric streams. SWAG will also reveal the heating/cooling
balance and chemistry of the gas in diverse environments and relates
those parameters to each cloud's evolutionary stage of star formation.

\end{document}